\documentclass[iop]{emulateapj}

\usepackage{verbatim}
\usepackage{graphicx}
\usepackage{placeins}

\slugcomment{ApJ, in press}
\shorttitle{The efficiency of star formation}
\shortauthors{Burkert \& Hartmann}

\begin{document}

\title{The Dependence of Star Formation Efficiency on Gas Surface Density}

\author{Andreas Burkert\altaffilmark{1,2} and Lee Hartmann\altaffilmark{3}} 

\altaffiltext{1}{University Observatory Munich, Scheinerstrasse 1, D-81679 Munich, Germany}
\altaffiltext{2}{Max-Planck-Fellow, Max-Planck-Institute for Extraterrestrial Physics, Giessenbachstrasse 1,
85758 Garching, Germany}
\altaffiltext{3}{Department of Astronomy, University of Michigan, 830 Dennison, 500 Church St., Ann Arbor,
MI 48109-1042, USA}

\email{burkert@usm.lmu.de, lhartm@umich.edu}

\newcommand\msun{\rm M_{\odot}}
\newcommand\lsun{\rm L_{\odot}}
\newcommand\msunyr{\rm M_{\odot}\,yr^{-1}}
\newcommand\be{\begin{equation}}
\newcommand\en{\end{equation}}
\newcommand\kms{\rm{\, km \, s^{-1}}}
\newcommand\K{\rm K}
\newcommand\etal{{\rm et al}.\ }
\newcommand\sd{\partial}
\newcommand\mpctwo{\msun \, {\rm pc^{-2}}}

\begin{abstract}
Studies by \citet{lada10} and \citet{heiderman10} have suggested that star formation mostly occurs
above a threshold in gas surface density $\Sigma$ of $\Sigma_c \sim 120 \, \msun \, {\rm pc^{-2}}$ ($A_K \sim 0.8$).
\citet{heiderman10} infer a threshold by combining low-mass star-forming regions, which show
a steep increase in the star formation rate per unit area $\Sigma_{SFR}$ with increasing
$\Sigma$, and massive cores forming luminous stars which show a linear relation.
We argue that these observations do not
require a particular density threshold. The steep dependence of $\Sigma_{SFR}$, approaching
unity at protostellar core densities, is a natural result of the increasing importance 
of self-gravity at high densities along with the corresponding decrease in evolutionary timescales.
The linear behavior of $\Sigma_{SFR}$ vs. $\Sigma$ in massive cores is consistent
with probing dense gas in gravitational collapse, forming stars
at a characteristic free-fall timescale given by the use of a particular molecular tracer. 
The low-mass and high-mass regions show different correlations 
between gas surface density and the area $A$ spanned at that density, with $A \sim \Sigma^{-3}$ for
low-mass regions and $A \sim \Sigma ^{-1}$ for the massive cores; this difference, along with
the use of differing techniques to measure gas surface density and star formation, 
suggests that connecting the low-mass regions with massive cores is problematic. 
We show that the approximately linear relationship between dense gas mass and stellar mass used by \citet{lada10} 
similarly does not demand a particular threshold for star formation, and
requires {\em continuing} formation of dense gas.  
Our results are consistent with molecular clouds forming by galactic hydrodynamic flows with subsequent gravitational collapse.
\end{abstract}


\section{Introduction}

Local star formation has long been known to be strongly enhanced 
in the densest regions of molecular clouds 
\citep[e.g.,][]{lada92, lada93, mizuno95, onishi98, johnstone04}.
The increasingly large and sensitive surveys of molecular clouds 
using a variety of techniques
have now made it possible to derive more quantitative relationships
between star formation and dense gas. In particular, using measurements
of infrared extinction, \citet[L10]{lada10} and \citet[H10]{heiderman10} 
proposed that there exists a steep decline in star formation surface density
$\Sigma_{SFR}$ and the ratio of $\Sigma_{SFR}$ to molecular gas surface density $\Sigma$
below a critical value of 
$\Sigma_c \sim 120 \, \msun \, {\rm pc}^{-2}$, 
approximately the same threshold \citet{onishi98} derived from C$^{18}$O 
observations of the Taurus molecular cloud.  
In addition, H10 and \citet{lada10,lada12}
inferred that star formation rates (SFRs) depend approximately linearly
on the amount of molecular gas above the threshold, qualitatively
consistent with findings from extragalactic studies \citep{gao04},
although with differing normalizations. Other extragalactic studies find a linear
relationship between the star formation of galaxies and their total molecular gas mass
\citep{bigiel11,krumholz12} which, combined with the correlation of
star formation and dense molecular gas mass might indicate a constant
ratio between the mass of dense molecular gas to diffuse molecular gas in galaxies.
On the other hand, \citep{gutermuth11} find no evidence for a surface density threshold
in their study of eight nearby star-forming clouds.

In this paper we investigate the existence of a star formation threshold density 
of $\Sigma_c \sim 120 \, \msun \, {\rm pc}^{-2}$. 
We show that the observations of steeply increasing star formation
with increasing gas surface density in low-mass star forming regions by H10 are consistent
with a continuous progression of the increasing importance of gravity and
decreasing evolutionary timescales with increasing density.  The impression of a threshold
in their data comes from combining results from low-density regions with those of
massive cores, which show a roughly linear dependence of the star formation rate on $\Sigma$.
However, we show that these two differing types of regions exhibit 
a differing dependence of the area $A$ at a given
surface density $\Sigma$, $A(\Sigma) \sim \Sigma^{-3}$ for low-mass regions, 
$A(\Sigma) \sim \Sigma^{-1}$ for massive cores, suggesting that the threshold
is the result of combining results from regions with very different physical conditions.
We suggest that the linear dependence of the star formation rate (SFR) per unit 
area and the surface density found by H10 in massive cores is a product of strongly-bound,
rapidly-collapsing regions with common free-fall times due to the use of a particular
molecular tracer.  Finally, we argue that the observations of L10 require
{\em continuing} formation of dense gas.  Both the development of power-law density
distributions and continuing formation of the densest star-forming gas are consistent 
with models of molecular cloud regions undergoing large-scale gravitational collapse 
\citep[e.g.,][]{hbb01,vs07,heitsch07,heitsch08a,heitsch08b,hennebelle08,banerjee09,zamora12}.

\section{Evidence for a surface density threshold?}

To examine the evidence for a threshold surface density in detail, we consider the
findings of H10.  These authors established roughly equally-spaced contour levels of 
(infrared) extinction in a sample of low-mass star-forming regions,
and then measured both the area and the number of protostellar (Class I and flat-spectrum) 
sources contained within each extinction or surface density contour.  
By focusing upon objects thought to be protostars, and thus a) very young and b) 
still accreting from their natal envelopes, H10 tried to minimize the 
conceptual difficulties which arise because the gas one observes at the present
epoch is {\em not} the gas that produced the stars.  

\begin{figure}
\includegraphics[width=0.45\textwidth]{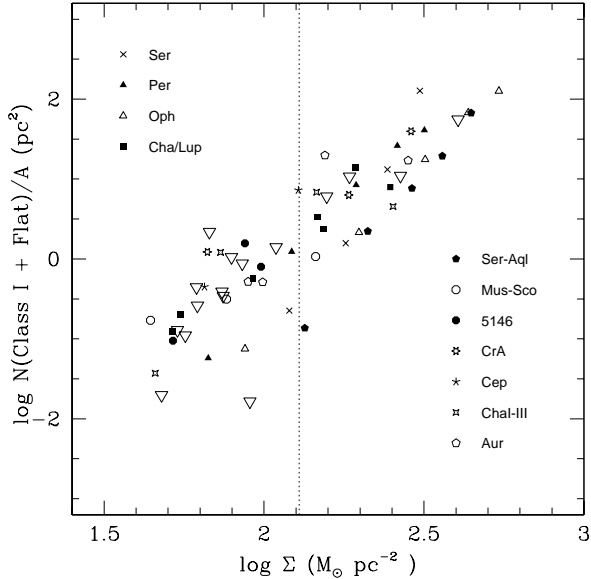}
\caption{The correlation of the number of protostars per unit area as a function of
$\Sigma$, using the data of H10 (their Table 2).  
Differing symbols indicate differing clouds, labeled by the identifiers used by H10.
Downward pointing large open triangles show ``upper limits'', regions
where no protostars were found, and so a value of one protostar was assigned for
plotting purposes, as in H10.   This is essentially the data plotted in Figure 8 of H10,
except that we have not divided by star-forming timescales,
have added the number of Class I and flat-spectrum sources together,
and did not include data from the HCN clumps (see text).
} 
\label{fig:firstbest}
\end{figure}

Figure \ref{fig:firstbest} shows the main results
of H10 for the low-mass star-forming regions.  We plot this as the number of
young stars $N$ per unit contour area $A$ vs. $\Sigma$; this is essentially comparable to
the star formation rate of H10. 
We also have added together the Class I and flat-spectrum
sources to improve the small number statistics).  The ratio $N/A$ is steeply dependent upon the 
molecular gas surface density, increasing by about three orders of
magnitude over a range of about one order of magnitude in $\Sigma$. 
H10 thus state that ``We identify this steep change in $\Sigma_{SFR}$ ... as a star-forming threshold 
$\Sigma_{th}$ between regions actively forming stars and those that are forming 
few or no low-mass stars.''

\begin{figure}
\includegraphics[width= 0.9\textwidth]{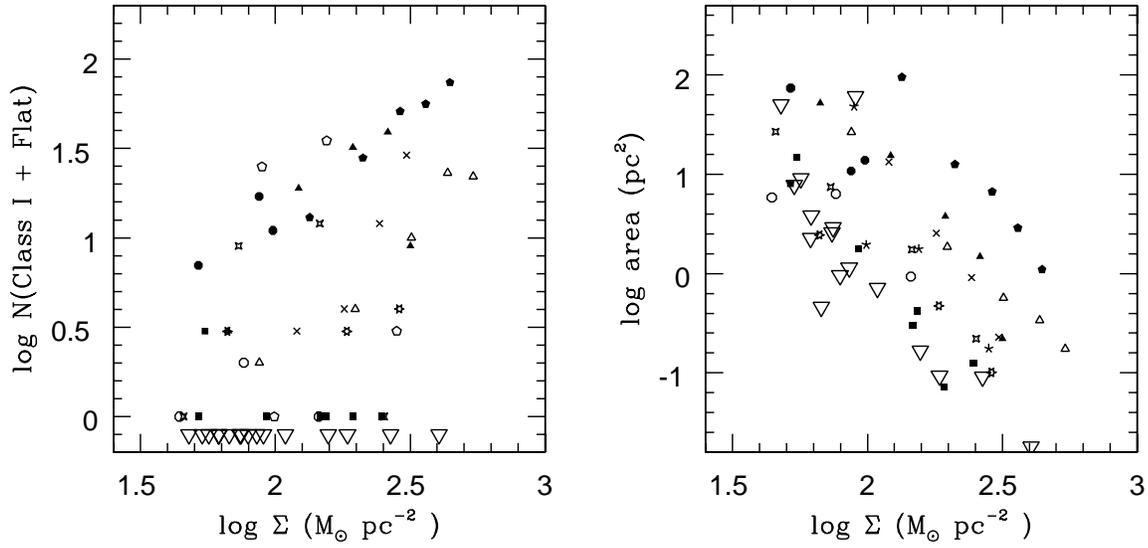}
\caption{Left: relation between the number of protostars within a contour
of area $A$ as a function of $\Sigma$.  The correlation is not strong,
though individual regions (Ser-Aql, pentagons; Oph, open triangles; Cep, crosses;
Per, filled triangles) show a trend of $N$ increasing with $\Sigma$.
The weak correlations are probably mostly due to the small number statistics,
as evidenced by the contours containing zero protostars (downward-pointing 
open triangles).  Right: area of each contour vs. $\Sigma$ of that contour.
There is a clear correlation; moreover, individual regions show similar
slopes with vertical displacements.  The results indicate that the correlation
seen in Figure \ref{fig:firstbest} is driven mainly by the $A(\Sigma)$ relation.
Symbols as in Figures 1 and 8.
}
\label{fig:twobest}
\end{figure}

While the $N/A$ vs. $\Sigma$ relation appears well-defined, it is in fact more of 
a reflection of the behavior of $A$ vs. $\Sigma$ than of the number of protostars.  
This can be seen in the left panel of Figure \ref{fig:twobest}, which shows a weak
correlation between the number of protostars in
a contour at a given $\Sigma$. The correlation shown in Figure \ref{fig:firstbest}
is driven by the area in a specific contour
$A$ vs. $\Sigma$ behavior, which is shown in the right panel of
Figure \ref{fig:twobest}.  Of course protostars are formed necessarily by high-density
gas; the lack of correlation in the left panel is probably mostly a result of small number
statistics, as even at high $\Sigma$ there exist regions
with no embedded protostar (downward-pointing triangles in Figure \ref{fig:twobest}).
In addition, one can see evidence for correlations between $N$ and $\Sigma$
for individual regions with reasonable stellar statistics (for example, Oph, Cha, and
Ser-Aql; see Figure \ref{fig:firstbest} for symbol identification).  Nevertheless, the
slope of $N/A$ vs. $\Sigma$ is basically determined by the slope of the $A(\Sigma)$
relation.

\begin{figure}
\includegraphics[width=0.4\textwidth]{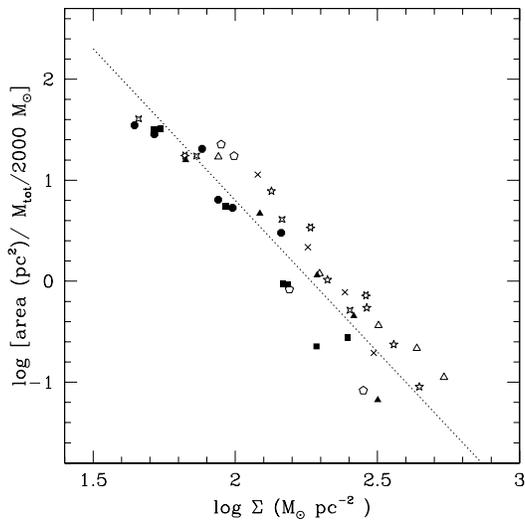}
\caption{
Areas at differing extinction contour levels, normalized by the scaled
total cloud mass above an extinction threshold $A_V = 2$ (from Table 1 in H10)
for the clouds in Table 2 of H10 with
more than one young star contained within the contour.  
The dotted line shows the least squares fit.
}
\label{fig:asigma.region}
\end{figure}

It is clear from the right panel of Figure \ref{fig:twobest} that a large
fraction of the scatter in the $A(\Sigma)$ relation
is due to the superposition of different regions (coded by different symbols), each of which
having intrinsically similar slopes but differing offsets.  This leads to the
plausible conjecture that more massive clouds have larger areas at a given surface density,
leading to correspondingly more stars per $\Sigma$ bin.
To investigate this possibility, in Figure \ref{fig:asigma.region} we plot the
area divided by a scaled total mass from Table 1 of H10.  The result is a greatly reduced
scatter around the trend, with the least-squares fit

\vspace{0.4cm}
\noindent
$\log A ({\rm pc^2})/M_{tot} (2000 \msun) = $
\be
(-3.24 \pm 0.18) \log (\Sigma/120 \mpctwo) + 0.62 \pm 0.05 \,.
\label{eq:asigscaled.region}
\en
Here we have chosen to normalize at the approximate surface density ``threshold''
$\Sigma_c = 120 $M$_{\odot}$pc$^{-2}$ of L10 and H10.\footnote{This is {\em not} the Larson (1981) scaling relation,
$\rho \propto r^{-1.1}$, which would imply $\Sigma \approx$~constant \citep[see discussion in]
[]{bp12}}. The power-law behavior $A \sim \Sigma^{-3}$ is reasonably consistent with that observed in
studies of the probability density function of column density \citep[e.g.,][]{K09,FR10},
although \citet{FR10} suggest that the slope may vary depending upon spatial resolution.

The local {\em volume} density $\rho$ is arguably at least as important, if not more
so, than $\Sigma$.
There is no unique way to convert a surface density to a volume density distribution without
using density diagnostics.  However, it is instructive to make the following guess.
We suppose that the area projected upon the sky $A$ at a given value of $\Sigma$ is
representative of that which would be seen from an orthogonal projection, at least on average.
In other words, we assume that the characteristic
average scale length along the line of sight is  $h \sim A^{1/2}$; thus
\be
\rho \sim \Sigma/A^{1/2} \,.
\en
For example, using equation \ref{eq:asigscaled.region}, the
value of the area of the contour at the critical surface density
is $A(\Sigma_c) \sim 4$pc$^2 (M_{tot}/2000$M$_{\odot})$;
this implies a median length scale for $M_{tot} \approx 2000 M_{\odot}$ of
$h \sim 2$~pc.  For $\Sigma_c = 120 \mpctwo = 2.5 \times 10^{-2} {\rm g cm^{-2}}$,
or a column density $N(H_2) = 5.4 \times 10^{21} {\rm cm^{-2}}$, the average
volume density is then $n(H_2) = 10^3 {\rm cm^{-3}} (M_{tot}/2000 M_{\odot})^{-0.5}$.

This analysis leads to an important point;
in proceeding to higher values of $\Sigma$, H10 is probing not only higher densities but
also {\em smaller masses and smaller volumes}.
Thus, the order of magnitude range in $\Sigma$ probed in Figure \ref{fig:firstbest}
corresponds to a range of two orders of magnitude in mass and three orders of magnitude in $A$; 
by the above argument,
this implies a range in $\rho$ of roughly a factor of 300, and thus a decrease
in the free-fall timescale by a factor of 17.
It is therefore not surprising that the number of young stars per
unit area, or the star formation rate per unit area as in H10, is a rapidly increasing
function of $\Sigma$ (see discussion in \S \ref{sec:thresholds}).

\begin{figure}
\includegraphics[width=0.45\textwidth]{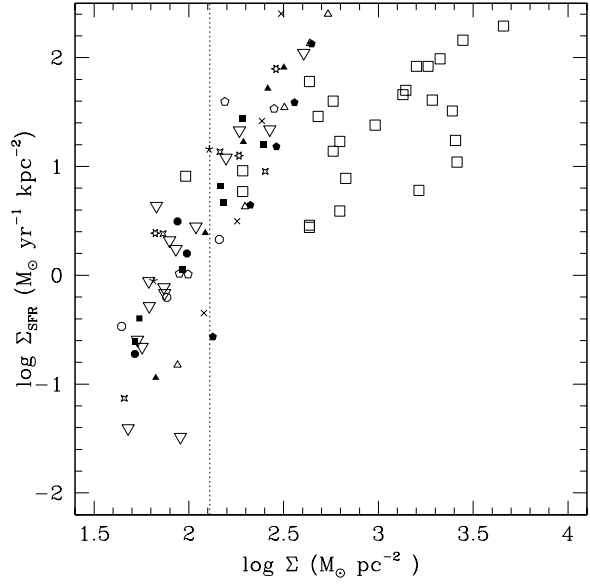}
\caption{The low-mass regions of Figure \ref{fig:firstbest}, plotted
now in terms of star formation rate per unit area, with the massive cores
studied by \citet{wu10} of luminosities greater than $\log L/\lsun \geq 4.5$,
using the calibration to star formation rate adopted by H10.
The low-mass star formation rates are slightly different than those of H10,
who used different lifetimes of 0.55 and 0.36 Myr for the Class I and flat-spectrum
sources, and used an average mass of 0.5 $\msun$; we use a lifetime of 0.5 Myr
and an average mass of 0.7 $\msun$ \citep{hh98}.
The dotted vertical line indicates the approximate location of the suggested
threshold.}
\label{fig:massivesfr}
\end{figure}

The impression of a threshold for star formation is enhanced in H10 by their
inclusion of an additional data set spanning higher surface densities than present
in their study of low-mass regions.  Analysing data from massive clumps with
surface densities measured from HCN and star formation rates SFR estimated from
infrared luminosities $L_{IR}$ from \citet{wu10}, H10 find a nearly linear relationship
between the SFR and $\Sigma$, spanning the range from $300 \lesssim \Sigma/\mpctwo \lesssim 3000$.
They interpret the intersection of this approximately linear massive clump branch
with the low-mass star formation branch
(Figure \ref{fig:massivesfr}) as a signature of a threshold at about $\Sigma_c = 120 \mpctwo$.

However, as H10 acknowledge,
there are concerns about combining results using two different methodologies, from
two very different types of regions (low-mass star-forming regions vs. massive cores
with $L_{IR} > 10^{4.5} \lsun$ and thus forming massive stars) (see also \S \ref{sec:discussion}). 
First of all, neither the low-mass nor the high-mass regions, by themselves, show a break;
it is only by combining the two that a ``break'' is seen.  
The results from the low-mass regions do not identify a particular value of $\Sigma_c$ but
instead extend smoothly into regions with $\Sigma > \Sigma_c$.  As a result, the intersection of the
linear fit of high-mass star forming clumps is not where the low-mass star formation branch ends,
but instead intersects the low-mass branch somewhere as one would expect if both branches are 
two independent correlations with different slopes. 

\begin{figure}
\includegraphics[width=0.9\textwidth]{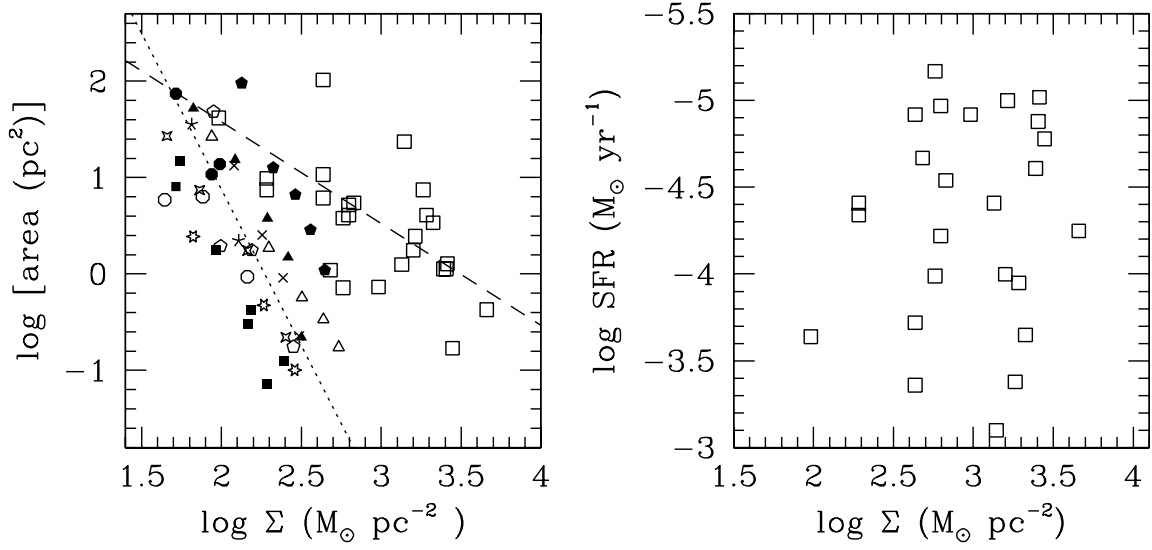}
\caption{Left: The $A(\Sigma)$ relations for both the low-mass regions
(see Figure \ref{fig:twobest}) and the
massive luminous core sample of \citet{wu10} (open squares).  
The dotted line is the fit given by equation (1); the dashed line is the approximate
fit to the massive cores of $A \sim \Sigma^{-1}$, indicating a constant core mass of
order $3000 \msun$ that is independent of $\Sigma$.  
Right: The massive cores show no correlation of star formation rate SFR with surface density,
so that the correlation of SFR/area vs. $\Sigma$ is driven by the $A(\Sigma)$ relation.
}
\label{fig:massivearea}
\end{figure}

One may also ask whether the physics of massive star formation is different, or whether 
the origin of two branches might simply result from the very different physical properties of the two types of regions?
An argument for the latter can be made by again looking at the behavior of $A(\Sigma)$.
To determine the areas fo the massive cores, we used the distances and
the HCN 1-0 FWHM sizes of \citet{wu10} in their Tables 1 and 6. 
(These are consistent with the surface densities and virial masses given in their Table 11.)
As shown in Figure \ref{fig:massivearea}, the $A(\Sigma)$ relation for the massive clumps
is not only displaced from that of the low-mass regions, but it has a different slope,
though with large scatter.  Again, as shown in the right-hand panel, the total star formation
rate is uncorrelated with $\Sigma$.  It makes physical sense that quantities per unit area
should be compared; nevertheless, in this data set the correlation is driven by $A(\Sigma)$, just as in the
low-mass case.  The difference in the two relations, with $A \sim \Sigma^{-3.2}$ for low-mass star-forming regions and
$A \sim \Sigma^{-1.1 \pm 0.2}$ for massive cores suggests that one is measuring differences
in physical conditions, not necessarily differences in the star formation process.

\begin{figure}
\includegraphics[width=0.45\textwidth]{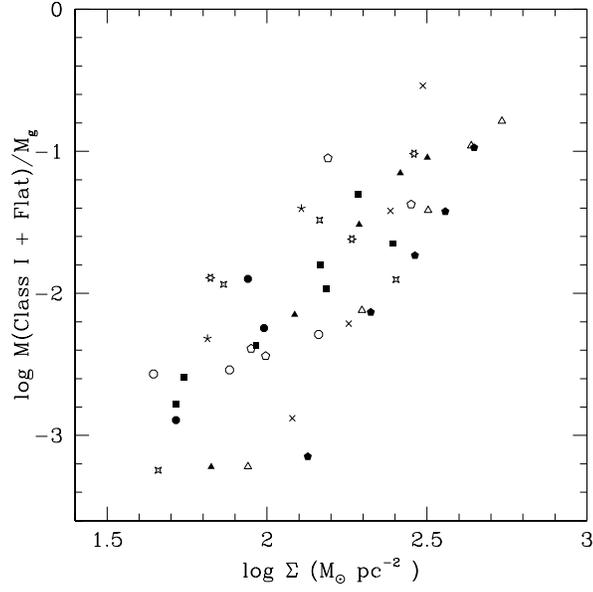}
\caption{The total protostellar mass (as estimated from the total
number of Class I and flat-spectrum sources, multiplied by an average mass
of $0.7 \msun$, and divided by the gas mass within the contour,
as a function of $\Sigma$.  Note that the slope is not significant
because the quantities in both axes depend upon $\Sigma$
(i.e., $M_g = \Sigma \times A(\Sigma)$).  However, it
is suggestive that, at high $\Sigma$, the ratio of stellar to gas mass
begins to exceed 10\% (see text)
}
\label{fig:lowmasseff}
\end{figure}

Assuming for the moment that high-mass star formation proceeds in a similar way as low-mass
star formation, how can we then reconcile the linear behavior of the high-mass star formation branch
with the strongly non-linear relationship, observed for the low-mass star forming regions?
The linear, high-mass star formation branch traces gas with molecular core densities of $\geq 10^5$ cm$^{-3}$.
To examine the implications for low-mass star formation we examine the 
star-gas correlation at these high densities.
In Figure \ref{fig:lowmasseff} we show the ratio of protostellar
masses to the mass of gas within the contour at $\Sigma$ \footnote{We caution that
this plot shows intrinsically correlated parameters, as the gas mass is derived by
multiplying the area of the contour by $\Sigma$, so the slope of the relation is
not significant.}.  At high surface densities the ``efficiency'' of star formation,
as measured by the ratio of dense gas mass to protostellar mass, exceeds 0.1. Interestingly, if we 
extrapolate
the correlation depicted in Figure \ref{fig:firstbest} to molecular core surface densities of 
$\Sigma \sim 1000 \, \msun \, {\rm pc}^{-2}$ which are similar to the massive core surface densities, 
the star formation efficiency approaches values of order unity. 
At this point the non-linear increase
of stellar mass versus dense gas mass with surface density must break
down as it is unlikely that more than 50\% of a core turns into stars \citep{federrath13}. If there
exists a maximum or typical efficiency for dense cores to form stars one would 
therefore naturally expect to observe a transition towards a more linear $N/A$ vs. $\Sigma$ relation.
As suggested above,
if in the massive cores the HCN tends to trace these particular, dense environments and thus also
a particular dynamical collapse time, a linear relation between the gas mass and star formation
rate might ensue.

This interpretation of the linear SFR vs.\ $\Sigma$ relation as simply observing
gas which is forming stars at maximum efficiency is further supported by the
fact that the use of infrared luminosity to determine the SFR for the HCN
clumps by \citet{wu10} is only possible as
reasonably massive stars have {\em already} formed in these regions.
If by using HCN a specific density range is being identified, and
thus a specific free-fall time, a linear relation between the star formation rate and
gas mass follows.
 
\section{Continuous molecular cloud evolution and gravity induced apparent thresholds}
\label{sec:thresholds}

\subsection{Basic Considerations}
\label{sec:overview}

H10 suggested that the rapid increase of $M_*/M_g$ for surface densities that lie within 
a factor of 3-4 around $\Sigma_c$ could be interpreted as a threshold for efficient star formation  
(\S 2).  Here we argue that this rapid increase will naturally occur in continuously-evolving clouds 
because: most molecular clouds form at modest densities and pressures;
the gas must become much more dense on its way to form a star; the evolutionary (gravitational
collapse) timescales rapidly become shorter as the gas becomes denser; and therefore
the contrast of low-density, slow evolution vs. high-density, fast evolution leads to the
impression of a threshold. 

Consider an estimate of the central pressure of a gas cloud 
$P_c = P_{ex} + \pi G \Sigma^2/2$, where $P_{ex}$ is the external
gas pressure and $\Sigma$ is the surface or column density
\citep{ee78}. This relation holds exactly for an infinite hydrostatic sheet 
and is roughly correct for many other geometries.  
Now, at the suggested critical surface density of 
$\Sigma_c \sim 120 \msun {\rm pc}^{-2}$, 
$P_c \sim 100 \times P_{ex}$ if we assume a typical local interstellar medium
(ISM) pressure $P_{ex}/k \sim 10^4 {\rm \, cm^{-3} \,K}$.

To achieve such ram pressures from flows at $10 \kms$,
typical of the turbulent ISM velocity dispersions in regular star-forming galaxies
\citep{dib06}, would require very large external flow densities of $\gtrsim 35 {\rm cm^{-3}}$.
In addition, to produce a region of this extinction, corresponding to hydrogen column density of
$N_H \sim 1.1 \times 10^{22} {\rm cm^{-2}}$, would require the converging or expanding flows to 
last for a timescale
\begin{equation}
\tau = \frac{\Sigma}{2 v_{\infty} n_{\infty}}
= 8.5 \times 10^7 {\rm yr} \left(\frac{10 \kms}{v_{\infty}}\right) \left(\frac{2 {\rm \,cm^{-3}}}{n_{\infty}}
\right) \,,
\end{equation}
which is an order of magnitude larger
expected from large-scale numerical simulations
of the dynamics of the turbulent ISM in disk galaxies \citep{dobbs12} that indicate
cloud formation timescales of order $5 \times 10^6 - 10^7$ yrs; again, one would require
high initial densities of $20 - 40 {\rm \,cm^{-3}}$ to build the cloud in reasonable timescales.
In summary, surface densities as high as $\Sigma_c$ are unlikely to be reached as a result
of the ram pressure in converging diffuse gas flows \citep{heitsch07,heitsch08a}.

Interestingly, most of the mass of molecular (CO) clouds in the solar neighborhood
lies at column densities corresponding to $A_V \sim 1-2$ \citep[e.g.,][]{goldsmith08}; i.e.,
there is an extended, lower-density molecular envelope, which would be at best slowly contracting 
due to its self-gravity, surrounding the dense, strongly self-gravitating gas. 
At $A_V \sim 2$, or a surface density
$\sim 40 \mpctwo$, the pressure due to self gravity would be
$P_c/k \gtrsim 6 \times 10^4 {\rm cm^{-3} \,K}$, only about an order of magnitude
larger than the typical ISM pressure.  This is indeed a natural result of 
forming molecular clouds (at least in the solar neighborhood) by interstellar
medium (ISM) flows with densities of $\gtrsim$ a few ${\rm cm^{-3}}$ at velocities
of order $10 \kms$, as simulated by \citep{bp99,audit05,hennebelle07,vs07, 
heitsch08a, heitsch08b}.  Moreover, such surface densities {\em can} be formed in the 
requisite timescales, especially
if the inflowing material has slightly higher densities than that 
of the average ISM \citep{dobbs12}.

Thus, while the majority of gas in molecular clouds is at (relatively) low densities,
and can be produced by external ISM ram pressures,
gas at or above $\Sigma_c$ is very likely to be generated and
bound mainly by gravity.  Independent support for this argument 
comes from analyses of the probability density functions (Npdf; the equivalent
of the $A(\Sigma)$ relations discussed in \S 2 ) of the column densities
of molecular clouds \citep{K09,K11,FR10,schneider12}.   The Npdfs appear lognormal at low densities,
which is generally interpreted as the result of turbulence-dominated flows
\citep{vs94,ostriker01,federrath08}, but they exhibit power-law tails at high column densities 
similar to $\Sigma_c$ in
star-forming clouds, as do the results of H10 (\S 2). 
\citet{bp11} showed that such power-law tails naturally arise when 
gravitationally-driven motion becomes more important than pure hydrodynamic turbulence,
consistent with our picture. $\Sigma_c$ might therefore mark those regions of molecular gas
that are contracting by self-gravity.

But $\Sigma_c$ is not a strict threshold. Although H10 find a rapid increase in 
star formation efficiency of about
three orders of magnitude near $\Sigma_c$, it actually takes place over a range of one order
of magnitude in $\Sigma$ which corresponds to two orders of magnitude in
``self-gravity pressure''.  Moreover, we argued previously that this range in
$\Sigma$ corresponds to a factor of $\sim 300$ in volume density; this corresponds to
a decrease in the free-fall time $\propto \rho^{-1/2}$ of a factor of 17.
Thus the combination
of strong self-gravity and rapid evolution will enhance the impression of a threshold.


\subsection{Self-gravitating sheet simulation}
\label{sec:sim}

To illustrate our argument in a more graphical way, we use some simple numerical simulations.
We consider a massive sheet of gas
which could have arisen as a result of colliding flows as in the simulations of
\citet{vs07, heitsch08a, heitsch08b}.  We
ignore turbulent motions and assume that the cloud is initially completely quiescent
but generates supersonic velocities via global and local gravitational collapse.
The setup is basically the same as that used in \citet{bh04}, where we considered
a uniform circular sheet.

The sheet is in hydrostatic equilibrium and in pressure equilibrium
with a constant surrounding pressure of $P/k = 5 \times 10^4$ K cm$^{-3}$. The initial conditions are the radius $R$ of the sheet,
its mass $M_{sheet}$, and the sound speed $c_s$ of the gas.
We assume an isothermal equation of state with the pressure $P = c_s^2 \rho$.
Here, we took an initial cloud radius of $R = 10$~pc, a sound speed of $c_s = 0.2 \kms$ and
calculated results for initial sheet masses ranging from
$M_{sheet} =10^3 \msun$ to $6 \times 10^3 \msun$. 
The calculations were performed with the SPH code outlined by \citet{bb97} and \citet{ba09}.

\begin{figure}
\includegraphics[width=0.45\textwidth]{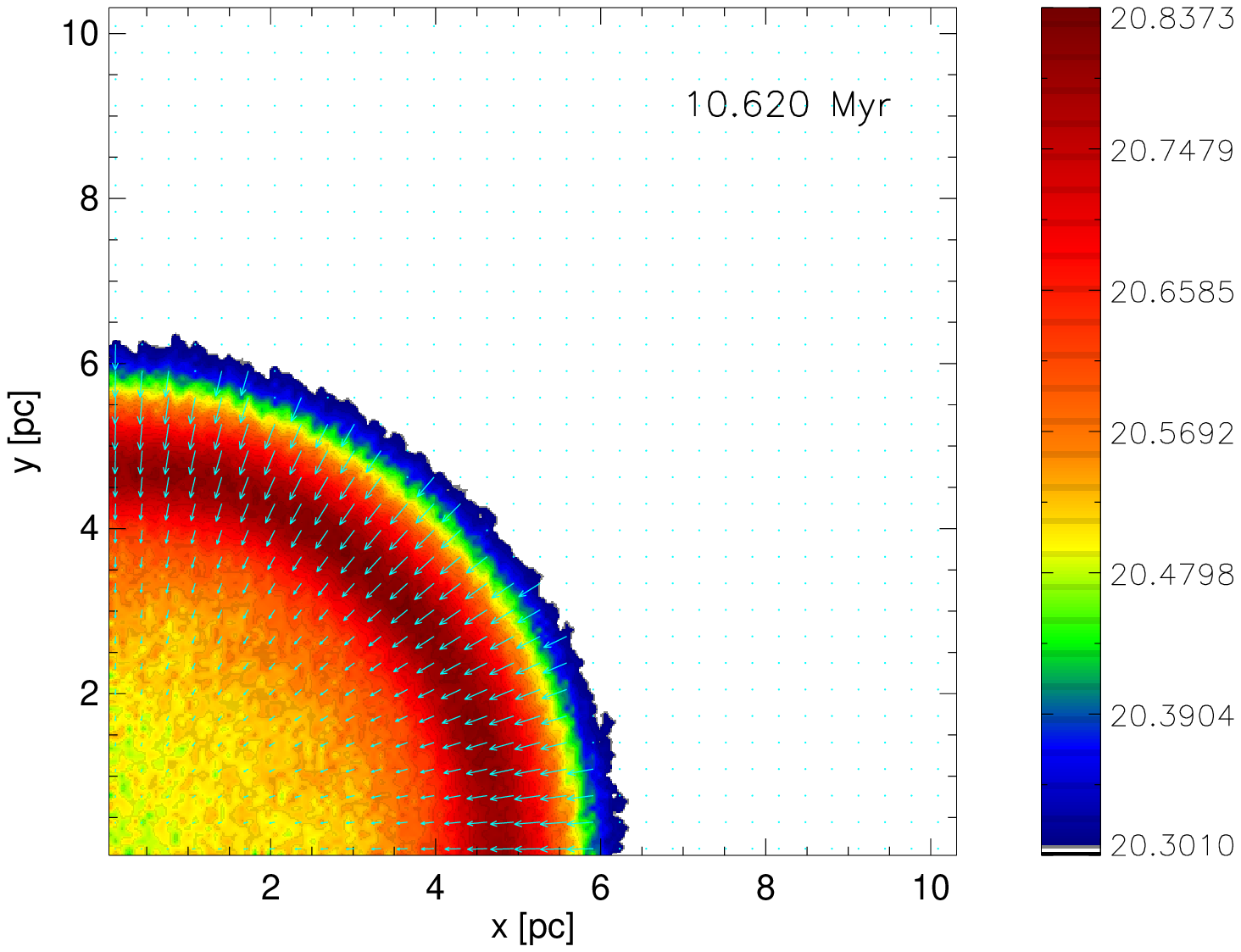}
\includegraphics[width=0.45\textwidth]{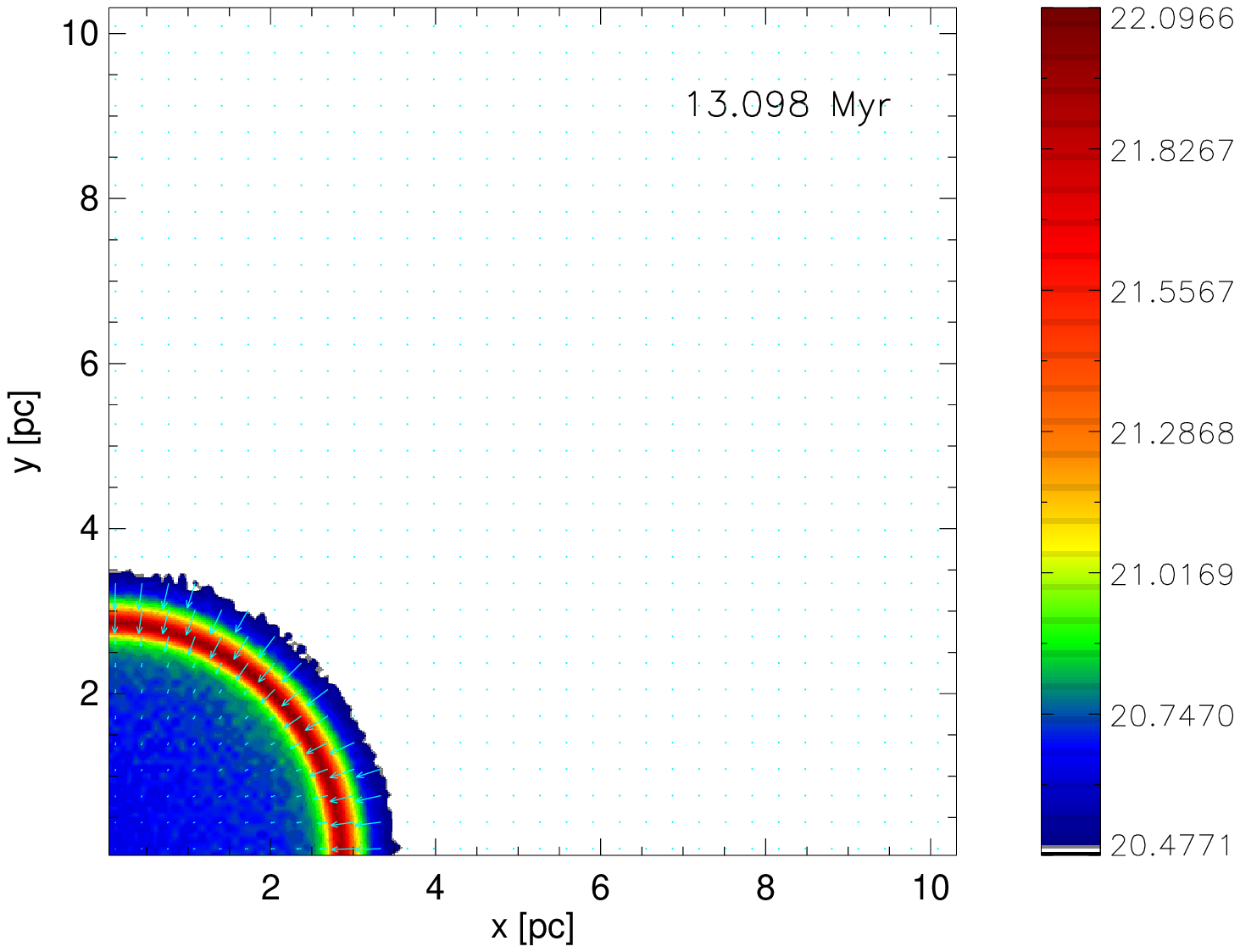}
\caption{Evolution of the collapsing sheet (standard model).
The surface density is plotted in logaritmic units of
$N(H_2)$ (see text).}
\label{fig:ring}
\end{figure}

Figure \ref{fig:ring} shows the surface density distribution of our standard model with $M_{sheet} = 10^3 \msun$
at two times.  With the adopted initial conditions,
a dense ring develops near the edge of the cloud, due to highly non-linear accelerations
of gravity \citep[e.g.,][]{bh04}.
In effect, we have made a circular and in this case artificially smooth, dense filament,
situated within a lower-density cloud.
The ring/filament continues to grow as the cloud (which has many Jeans masses)
globally contracts, accumulating mass and becoming denser as time proceeds.

\begin{figure}
\includegraphics[width=0.5\textwidth]{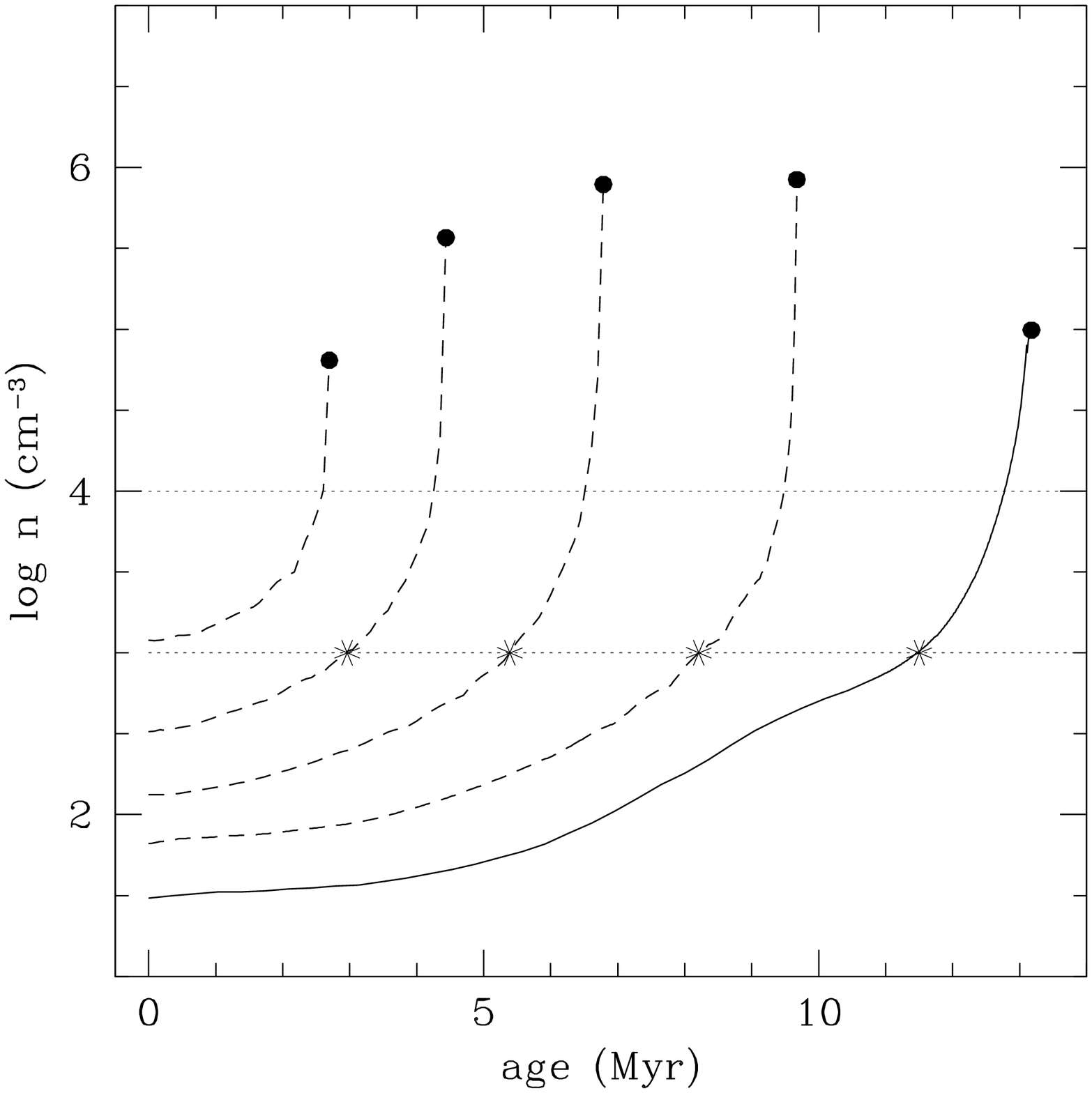}
\includegraphics[width=0.5\textwidth]{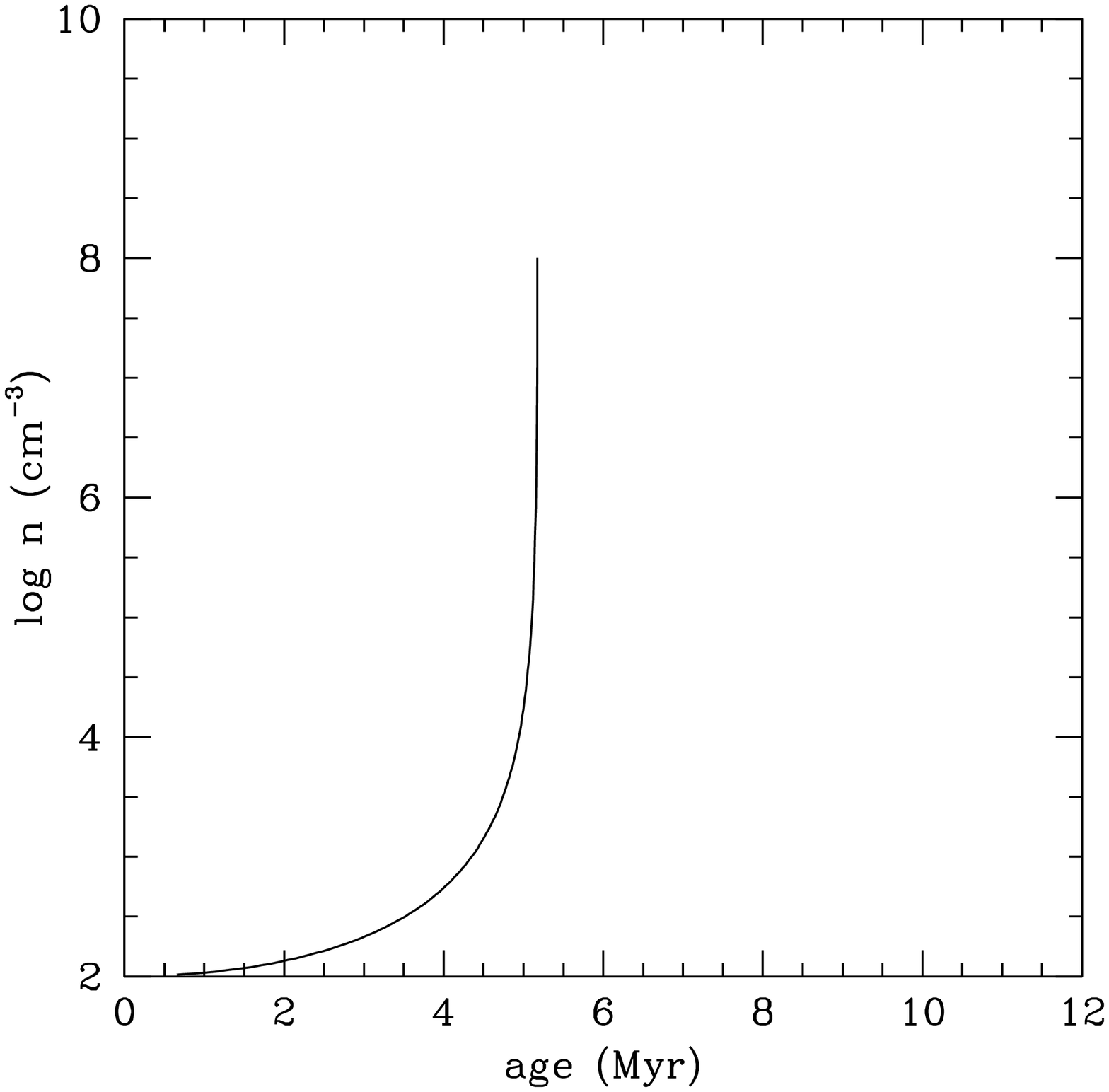}
\caption{Left: 
Evolution of mean ring densities for the standard model with 
$M_{sheet} = 10^3 \msun$ (solid curve) and simulations with initial masses
of $1.5 \times 10^3$, $2 \times 10^3$, $3 \times
10^3$ and $6 \times 10^3 \msun$.  The filled points show the time when 
collapsing fragments appear in the ring. Stars show the points in time at which
the rings in the sheet achieve densities where they are expected
to shield CO and become CO filaments (see text).
Right: evolution of density for the uniform free-falling sphere.}
\label{fig:circfrag}
\end{figure}

Figure \ref{fig:circfrag} (left) shows the evolution of the mean ring volume densities
for the standard model (solid line) as well as for additional models with differing initial masses
and thus surface densities.
The density evolution can be characterized by two phases, ``slow'' and ``fast'',
a distinction which is clearest for the two lowest-mass models.  The upper limit
to the ``slow'' timescale is the collapse time 
$t_{coll}$ for the sheet as a whole \citep{bh04},
\be
t_{coll} \approx \left ( {R \over \pi G \Sigma} \right )^{1/2}\,
\approx (G \langle \rho \rangle)^{-1/2}\,.
\label{eq:bh04time}
\en
In other words, $t_{coll}$ is the global free-fall time of the sheet characterized by
$\langle \rho \rangle$, the mean density over the spherical volume enclosed by $R$.
Equation (\ref{eq:bh04time}) yields 
a global collapse time of 15 Myr for $M_{sheet} = 10^3 \msun$. In this case the
sheet contracts considerably before the ring (filament) density runs away. Because
the sheet volume density scales as $\Sigma^2$, the timescale for collapse of the ring becomes
much faster with increasing surface density.

Figure \ref{fig:circfrag} (left) demonstrates our basic explanation of apparent thresholds
for star formation.  There are extended timescales over which the cloud remains at low column densities.
These timescales are given by the global sheet collapse timescale that depends on the
initial surface density.
Our filaments reach column densities sufficient to shield CO of $A_V \sim 1$ when
they achieve volume densities $\sim 10^3 {\rm cm^{-3}}$, indicated by the stars. At that time
the sheets are expected to convert into CO clouds.\footnote{This approximation for 
CO formation was similarly used by \citet{heitsch08b};
\cite{clark12} found that a somewhat higher extinction level should be adopted, but
that the general treatment is reasonable.  In our case the extinction is probably underestimated
given the sheet geometry we use; foreground and background gas would add additional shielding;
Conversely, this would mean that CO forms at lower densities.}
Evolution now becomes particularly rapid, with 
phases of $\sim 2$~Myr in the ``CO cloud phase'' to $\ll 1$~Myr above $10^4 {\rm cm^{-3}}$ until 
enough material has accumulated and pressure forces cannot support the ring anymore. It
becomes gravitationally unstable and collapses onto itself, leading to 
runaway growth in density that is much faster than the growth in mean density
in the sheet and resulting in two gas phases, a diffuse gaseous component 
(the central regions in Figure 6) and an
embedded dense collapsing filament that now can fragment and form stars. The collapse occurs
essentially at the free-fall time of the {\em local} density, very similar
to the evolution under pure free-fall that is shown in the right panel of Figure \ref{fig:circfrag}. 

We emphasize that in all cases the density evolution is continuous; there is no
particular density threshold at which one can say star formation will or will not occur
in the future.

Of course these simulations are also highly unrealistic, in that clouds will have multiple sites of
star formation; but one might view Figures \ref{fig:ring} and \ref{fig:circfrag}
as representing the evolution of differing parts of the cloud or even differing local
patches, with the same basic result.  Indeed, even the limiting case of the uniform
pressureless sphere collapse exhibits the aspect of slow initial growth and localised fast
runaway collapse.

Note that in our simulations, the entire cloud is gravitationally bound and
collapsing.  However, even if we do not support the low-density medium via e.g. turbulence or 
magnetic fields, a broad range of evolutionary timescales develops due to ring formation
\citep[see also][]{pon11}. 

\section{Efficiency of star formation in dense gaseous environments above $\Sigma_c$}

L10 examined the ratio of the gas mass above a given level of extinction 
(or surface density) $M_{dense}$ to the number of young stars $N$ in several 
nearby star-forming regions.  They searched for the value of $\Sigma$ 
which would most nearly result in a linear relation between $M_{dense}$ and $N$ 
for the various clouds; the resulting critical value was found to be
$\Sigma_c = 116 \pm 25 \, \msun \, {\rm pc}^{-2}$.  While L10 identified this
as a threshold for star formation, they did note that there was a factor of nearly two
in $\Sigma$ for which the dispersion in $M_{dense}/N$ among the various clouds was
minimal.

\begin{figure}
\includegraphics[width=0.5\textwidth]{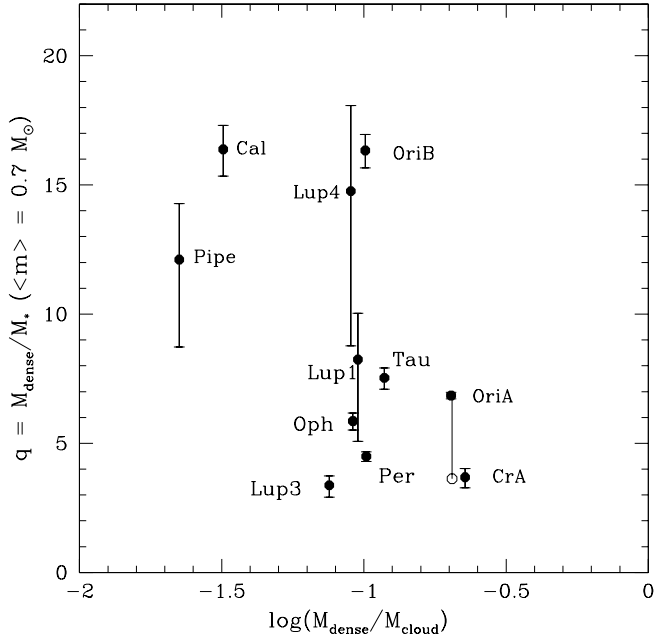}
\caption{The ratio $q$ of gas mass $M_{dense}$ above $\Sigma_c$ to the estimated
mass of young stars $M_*$, plotted as a function of the ratio of 
$M_{dense}$ to the total cloud mass, using data from Table 2 of L10.
A mean mass of $0.7 \msun$ per star has been used to convert
the number of stars to a total stellar mass.  Errorbars reflect {\em only} counting
statistics for the number of stars.  The L10 results for Orion A (solid dot)
have been adjusted to take into account more complete surveys of the
stellar population (open circle; see text).  There is an indication
that $q$ decreases as the fraction of dense gas mass increases.
}
\label{fig:qdat}
\end{figure}

As discussed earlier, we do not find any evidence that the value of $\Sigma_c$,
quoted by L10 is special in separating efficient from inefficient star formation.
In fact, the analyses of L10 is somewhat misleading. 
Note that in their Figure 2 they show the ratio of the number of young stellar objects N(YSO)
to total gas mass versus total gas mass
and in their Figure 4 they show N(YSO) versus dense gas mass. Figure 4 shows a good correlation
and Figure 2 shows no correlation. If however they would have plotted Figure 2 like Figure 4,
i.e. N(YSO) versus total, mass they would also find a good correlation (Krumholz, private communication).
The data presented in L10 however provides an interesting basis in order to investigate
how star formation occurs in a dense gaseous environment,  in this case regions with
$\Sigma_c \geq 116 \pm 25 \, \msun \, {\rm pc}^{-2}$. Let us therefore look in more details at the data.
In Figure \ref{fig:qdat} we plot the data from Table 2 of L10 
in the form of the ratio of the dense gas mass to stellar mass 
$M_{dense}/M_* = q$, vs.\ the ratio of the total gas mass $M_{tot}$ to $M_{dense}$,
again using an average mass per star of $0.7 \msun$.
The errorbars in Figure \ref{fig:qdat} indicate {\it only} counting statistics; the true uncertainties
are much larger.  For example, L10 cite a total of $\sim 2900$ young stars in
Orion A; however, \citet{hh98} estimated that there are $\sim 2200$ stars just within
15 arcmin of the center of the Orion Nebula, based on optical and deep near-infrared
surveys.  \citet{getman05} found $\sim 1400$ members within the 17' $\times$ 17'
square field of the COUP deep X-ray survey; this corresponds reasonably well to
the model of \citet{hh98} which predicted $\sim 1100$ members in the same area.
\citet{rebull00} found another $\sim 1600$ likely members in fields ``flanking'' the Orion
Nebula Cluster.  Finally, results based on {\it Spitzer Space Telescope} and optical spectroscopic
surveys indicate another $\sim 1600$ members in the region south of $-6^{\circ}$ (the ``extended''
L1641 region) \citep{2012ApJ...752...59H}.  The resulting total of $\sim 5400$ likely members is roughly
a factor of two larger than L10 cite for Orion A; the resulting $q$ is plotted as an open
circle in Figure \ref{fig:qdat}.  Similarly, it is likely that the
stellar population of Orion B is underestimated by an even greater factor, as the nebulosity and
high extinction makes even {\em Spitzer} surveys highly incomplete 
(S.T. Megeath, personal communication), and the region has not been surveyed as extensively
in X-ray and optical spectroscopic surveys which can find non-infrared excess sources
(weak T Tauri stars).  Finally, the above discussion deals only with the stellar population,
and not any potential errors is gas mass estimation. 

Discounting the Lupus 4 region for poor stellar counting statistics, 
and noting that the stellar population of Orion B is almost certainly understimated,
the data suggest that the least-active regions with large values of $q$ 
have low fractions of dense gas, while in the most active regions the fraction of dense 
gas is high, which may be an indication of cloud evolution.  For the most active 
and probably evolved star-forming regions, (apart from Orion B) the range in the gas to stellar mass is 
$8 \gtrsim q \gtrsim 3$. Note that here we focus on young star forming regions with the stars 
still being embedded in their cold molecular environment. Once stellar feedback becomes efficient,
q will quickly drop to small values.

An important implication can be derived from these results.  As all the young stars 
have not formed at the same time (indeed, some of them are still-accreting
protostars), $M_*$ must be increasing with time; therefore
the mass of dense gas $M_{dense}$ {\em must also increase in time}
to keep $q$ above unity and within a modest range.  
Including dispersal of gas by stellar
energy input only increases the need for additional dense gas formation.
Thus, the L10 results imply that star-forming
clouds are not isolated regions of mass but are continuously collecting gas from their
environment
\citep{dobbs11a,dobbs11b,dobbs12}.

\subsection{Star formation with constant dense gas to stellar mass ratio}

To illustrate further implications of the L10 data, we consider a simple analytic
model. Let us assume that a dense molecular gas component is generated at a rate $\dot{M}_{in}$.
Suppose also that a fraction $\epsilon$ of the dense gas turns into stars on its local free-fall 
timescale $\tau_{ff}$
\be
{d M_{dense} \over dt} = \dot{M}_{in} - {d M_* \over dt}\,,
\label{eq:mgevol}
\en
\be
{d M_* \over dt} = {\epsilon \over \tau_{ff}} M_{dense}\,,
\label{eq:msevol}
\en
\noindent where we take $\epsilon$ and $\tau_{ff}$ to be constants. 
Typical values of $\epsilon$ are $\epsilon \approx 0.1$ (L10).
Now, suppose the suggestion of L10 of a linear relation between the amount of dense gas
and the stellar population holds exactly, such that
at any given time $t$ the ratio $q=M_{dense}/M_*$ is constant.
We then find
\begin{equation}
M_{dense}(t)= M_{dense,0} \, \exp(t/t_0) \,,
\label{eq:mg}
\end{equation}
where $M_{dense,0} = M_{dense}(t=0)$ is the initial dense gas mass when star formation starts
and $t_0$ is the exponential growth timescale of the dense gas mass,
\begin{equation}
t_0 = \frac{\tau_{ff}}{q \epsilon}\,.
\label{eq:tgrowth}
\end{equation}
Inserting equation (\ref{eq:mg}) into equation (\ref{eq:mgevol}) we find
\begin{equation}
\dot{M}_{in} = \epsilon (q+1) \frac{M_{dense,0}}{\tau_{ff}} \exp \left(\frac{t}{t_0} \right)
\label{eq:min}
\end{equation}
Thus, if $q$ were absolutely constant, there would need to be
an {\em exponential infall and increase of dense gas mass with time}.

This exponential growth model raises some concerns.  Obviously
$M_{dense}$ cannot increase exponentially forever. In addition, as the fraction of dense
gas is observed to be always small compared to the diffuse gas mass and as the
efficiency of star formation is generally only a few percent of the total cloud  
mass \citep[e.g.,][]{evans09,krumholz12,federrath12}, exponential growth
would imply a very delicate timing for dispersal 
of the whole cloud. Stellar feedback would have to destroy the cloud before 
a large fraction of its diffuse gas has been converted to dense gas. Finally,
it is not clear that simulations provide support for 
continued exponential growth of the dense gas mass \citep[e.g.,][see \S \ref{sec:sim}]
{vs09}.  To address this problem we develop a second analytic model which assumes 
linear rather than exponential growth in the dense gas mass with time. 
Adopting time units $\tau = t/\tau_{ff}$ and assuming a constant $\dot{M}_{in}$, the solutions of
equations (\ref{eq:mgevol}) and (\ref{eq:msevol}) now are
\be
M_{dense}(\tau) = M_{dense,0} \, e^{-\epsilon \tau} + \dot{M}_{in} \, \tau_{ff} (1 \, - \, e^{-\epsilon \tau})/\epsilon,
\label{eq:mglinear}
\en
$M_*(\tau) = M_{dense,0}(\tau =0) \, (1 - e^{-\epsilon \tau})$
\be
+ \dot{M}_{in} \, \tau_{ff} \,[\tau \, - \, (1 - e^{-\epsilon \tau})/\epsilon] \,.
\label{eq:mstarlinear}
\en

\begin{figure}
\includegraphics[width=1.\textwidth]{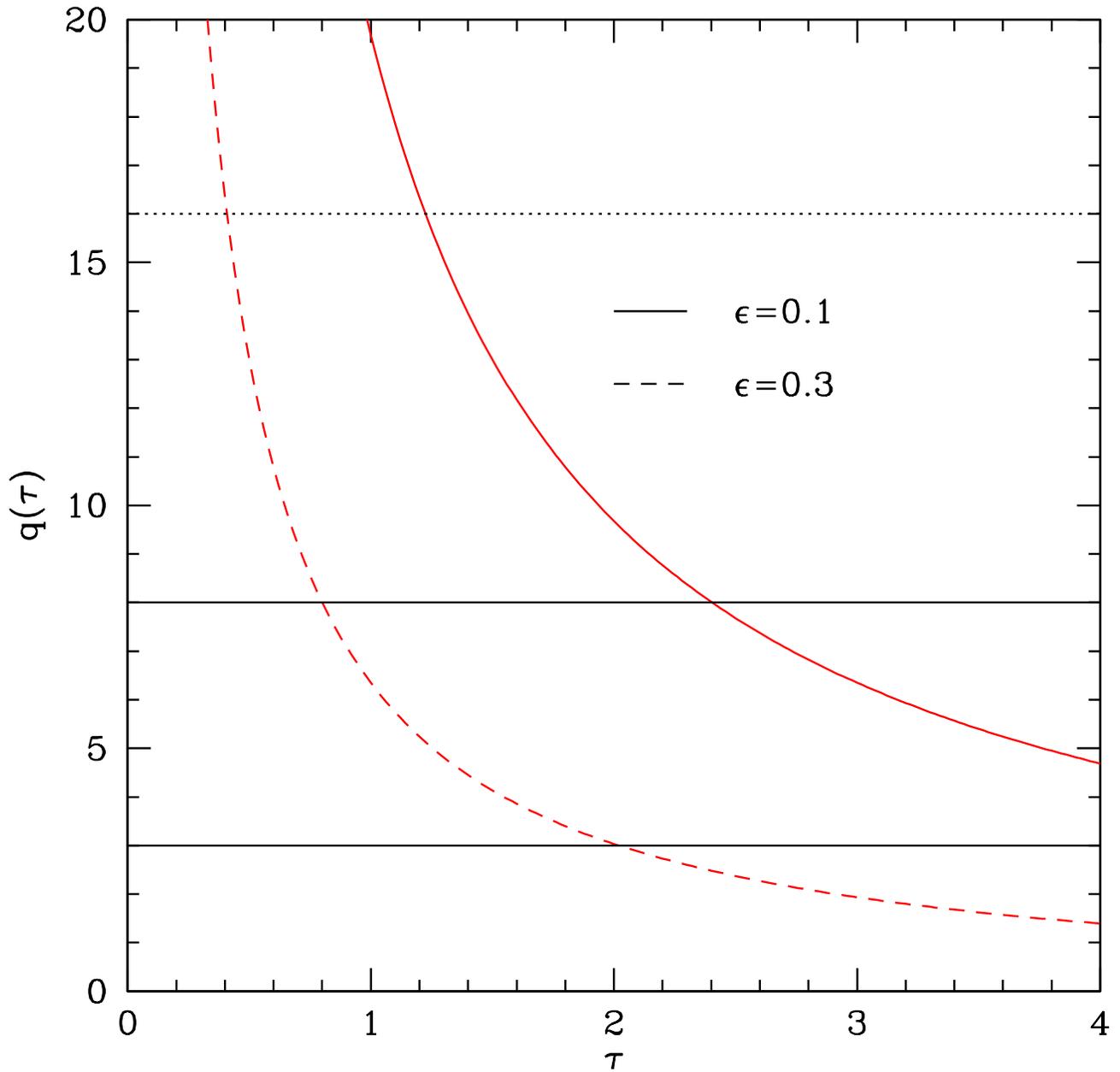}
\caption{Results for the solution of equations \ref{eq:mglinear} and
\ref{eq:mstarlinear}, and adopting $M_{dense,0} = 0$
for two values of $\epsilon$.  The dotted horizontal line indicates
the observed upper $q$, while the two solid horizontal lines indicate the empirical
limits for the ``evolved clouds'' (see text).}
\label{fig:linear}
\end{figure}

Figure \ref{fig:linear} shows results, assuming $M_{dense,0}=0$ for efficiencies of
0.1 and 0.3, respectively. 
For values of $\epsilon \approx 0.1$ that are consistent with the observations of L10,
$q$ remains in the observed range over $4 \tau_{ff}$.

\subsection{Star formation above a critical volume density}

L10 suggested that the critical surface density might correspond to a critical {\em volume}
density $n_c = 10^4$ cm$^{-3}$, implying 
$\tau_{ff} \sim 0.35$~Myr.  
(Note that our characterization of $\Sigma_c$ and  $n_c$ as ``critical'' simply refers to the
numerical values adopted by L10 rather than signifying true thresholds for star formation.)
Thus, with linear growth of the dense gas mass,
Figure \ref{fig:linear} indicates that even for star formation
efficiencies of $\epsilon = 0.1$, the observed
range of $q$ can only be maintained for less than 1-1.5 Myr, considerably shorter than
typical estimates of nearby star-forming regions within molecular clouds which are of order $2-4 \times 10^6$ yrs
(Hartmann \etal 2001; Hartmann 2001, 2003); L10 use $\sim 2$~Myr. 
However, there are reasons to question this estimate of the critical volume density.
If we use the data of H10 and
estimate the path length along the line of sight to be $l \sim A^{1/2}$ and thus
$<\rho> \sim \Sigma/A^{1/2}$, we find
a narrow range of estimated mean densities at $\Sigma_c$ between $n_c \sim 500 {\rm cm^{-3}}$
and $\sim 2.6 \times 10^3 {\rm cm^{-3}}$. Taking a mean value for the critical volume
density for all clouds of
$n_c \sim 10^3$~$cm^{-3}$  implies a free-fall time of $\sim$ 1 Myr. This makes
it much easier to explain the range of $q$ found by L10 as $q$ now would lie in the observed
regime for $4 \times 10^6$ yrs which are reasonable star formation timescales of 
nearby star-forming regions (Hartmann \etal 2001; Hartmann 2001, 2003).
As a consistency check, the areas at $\Sigma_c$ of one to several
pc$^2$ are also much larger than those typical of low mass pre-stellar cores of densities
above $\sim 10^4$~cm$^{-3}$ \citep[e.g.,][]{myersbenson83,enoch08}.

\begin{figure}
\includegraphics[width=1.\textwidth]{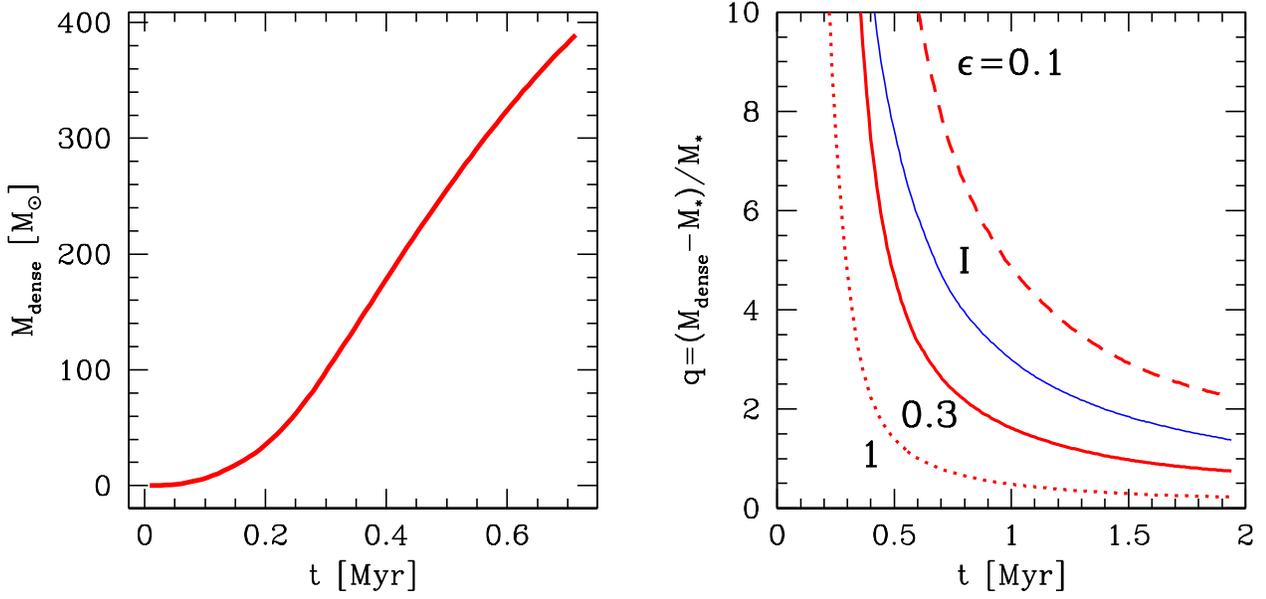}
\caption{ Inferred ratios of stars to dense ($n > 10^3 \, {\rm cm^{-3}}$) gas, using the standard model simulation and various assumptions
about star formation rates.  On the left, the evolution of dense gas with a time offset.
In the right panel, we show $q = M_{g}/M_*$ as a function of time assuming
$M_{g} = M_{dense} - M_*$, where $M_{dense}$ is the total gas as in the left-hand panel,
and $M_*$ the mass in the circular filament above that critical density
0.35 Myr earlier, times an efficiency factor $\epsilon$.
The dotted curve assumes $\epsilon = 1$, the solid red curve $\epsilon = 0.3$, and the
dashed curve $\epsilon = 0.1$.  The blue solid curve, labeled $I$, shows the situation assuming
that after every 0.35 Myr a new, independent dense gas region forms
that does exactly the same thing as the $\epsilon = 0.3$ case, just starting later (see text).
}
\label{fig:qsim}
\end{figure}

We can also use the numerical simulations to check the arguments constraining $q$. 
The left-hand panel in Figure \ref{fig:qsim} shows the growth of mass 
with densities above $10^3 {\rm cm^{-3}}$ for the standard model.
One observes that the growth of dense gas is exponential for the first $\sim 0.2$~Myr, but then becomes
roughly linear.  If we were to keep $q$ strictly constant, the cloud would
have to be dispersed at the end of the short epoch of exponential growth which is not in agreement
with observed star formation timescales. To illustrate the implications
further, in the right panel of Figure \ref{fig:qsim} 
we show $q = M_{dense}/M_*$ as a function of time for the standard model
with $M_*$ the mass in the circular filament above 
$n(H_2) = 10^3 {\rm cm^{-3}}$, but at a free-fall time of
1 Myr earlier, times an efficiency factor $\epsilon$. 
In other words, here we assume again
that $\epsilon M_{dense}$ turns into stars after one free fall timescale.
The dotted curve assumes $\epsilon = 1$, the solid red curve $\epsilon = 0.3$, and the
dashed curve $\epsilon = 0.1$.  

Figure \ref{fig:qsim} (right panel) reinforces the analytic results. For efficiencies of order 10\%
$q$ will lie in the range found by L10 for 4 free fall timescales.
If, in addition, the freefall timescale is of order 1 Myr, as expected
for critical densities of $n(H_2) = 10^3 {\rm cm^{-3}}$, q would remain in the observed
regime for the typical lifetimes of nearby star-forming
regions.  The simple analytic model actually does a good job of reproducing these
combined numerical-``post processing'' results in Figure \ref{fig:qsim} (for example,
it predicts that for $\epsilon = 0.1$, $q$ drops below $\sim 3$ at $4 \tau_{ff} \sim
4$~Myr).

Dense gas formation is probably not monolithic, but instead is the result of differing regions collapsing at
differing times.  The blue solid curve, labeled $I$, shows the situation if
we assume that after every $\tau_{ff}=1$Myr a new, independent dense gas region forms
that does exactly the same thing as the $\epsilon = 0.3$ case, just starting later.
So e.g. after $2$ Myr we have in total 3 star forming regions: one is 2 Myr old,
one is 1 Myr old and just beginning to form stars, and one is just starting to generate
dense gas but is not forming stars yet.  This spreading of the onset of star formation helps
in maintaining $q$ within the observed levels, leading to values of $q$ that are somewhat
larger than the values expected for a single star forming region with $\epsilon = 0.3$
(solid red curve).
(Note that we are not concerned with whether $q$ declines monotonically with time,
as this depends upon the coordination in time of star formation in separate regions.)

Finally, the behavior of $q$ implies something about the efficiency of turning gas into
stars.  Sometimes star formation sites are characterized by the ``star formation
efficiency per free-fall time'' $\eta_{ff}=SFR \times \tau_{ff}/M_{dense}$ \citep{krumholz12,federrath12}. 
Considering only the gas mass above $\Sigma_c$, L10 find a low value
$\eta_{ff} \approx 0.02$ which might at first indicate a surprisingly low value
of star formation efficiency. Note however that this result depends upon their assumption of 
$n_c = 10^4 {\rm cm}^{-3}$.  If instead $n_c = 10^3 {\rm cm}^{-3}$,
then $\eta_{ff} \sim 0.06$.  
In addition, the concept of $\eta_{ff}$ is complicated, not only because real clouds exhibit
a wide range of densities and thus regions with very different free-fall timescales \citep{federrath12},
but also because it depends upon the assumption of a quasi-steady
state.  For example, if the gas mass grows exponentially as in the model which provides
strictly constant $q$,
\be
\eta_{ff} = (q t/\tau_{ff} )^{-1}\,.
\label{eq:etaeff}
\en
Note that $\eta_{ff}$, despite its name, does not depend 
explicitly on the star formation efficiency
$\epsilon$, that is the fraction of dense gas that turns
into stars in $\tau_{ff}$. A low value of $\eta_{ff}$ is therefore not in conflict with theoretical
models that would predict higher values of $\epsilon$. Given $q$, $\eta_{ff}$ is instead
a measure of the lifetime $t$ of the star forming region in units of its free fall time
that is being continuously fed by infall from its diffuse molecular envelope.
Even if the growth is not strictly exponential, there is still a tendency to underestimate
$\eta_{ff}$, as well as significant uncertainty as to what the efficiencies were in the past when
the first stars formed \citep[e.g.,][]{vs09}. 

\section{Discussion and Conclusions}
\label{sec:discussion}

We have shown that the observations of H10 and L10
do not require density thresholds for star formation; all that is required is 
rapid gravitational collapse at high densities, coupled with the
presence of a much lower-density molecular cloud formed via plausible interstellar
medium flows.  This ``external'' low-density molecular material feeds the dense
regions.  The low-density gas can also be gravitationally collapsing and still
provide a more slowly-evolving structure with more mass than in the dense regions,
especially if the global cloud geometry is far from spherical.
If the flows into the dense regions are driven by gravitational acceleration
the increase of dense gas mass with time to keep the ratio of
dense gas to stars relatively constant naturally occurs.  The resulting
picture is consistent with simulations of dynamically-evolving, gravitationally-collapsing
star-forming molecular clouds.  Density diagnostics (e.g. \citet{brunt10}) are needed to help
translate surface densities into volume densities, which then will constrain
the free-fall times and thus the efficiences of star formation at modest
surface densities.  

The data of H10, combined with the massive core results from \citet{wu10}
suggest that at high $\Sigma$ the efficiency of
converting the mass of gas into stars appears to approach unity (see also
\citet{masiunas12,federrath13}).
We suggest that the linear relationship between SFR and $\Sigma$ then arises
because this dense gas is in gravitational collapse and efficiencies above unity are
unlikely due to stellar feedback.  The survival of this linear
relationship in extragalactic, beam-diluted observations suggests an
approximate common scaling of the total mass of gas with gas mass above
$\Sigma_c$; an open question is whether this is a result, or can be derived from,
the apparent ``power-law'' distribution of areas as a function of density. 
The origin of the $A(\Sigma)$ relation is unclear and has to be related to the
universal density structure of clouds; the transition between the log-normal
form of the Npdf at low column densities to the power law behavior we have found in
the data of H10 and seen in other investigations
is likely the product of gravitational collapse \citep{bp11}.

Along the way we have noted some of the conceptual difficulties in deriving
density thresholds for and efficiencies of star formation.  The dense gas we
see at the present epoch is not the gas that formed, or is forming the young
stars currently present. Can we then assume a steady state to interpret the observations
and linking past star formation to present cloud structure?
In the case of galactic molecular clouds, we have argued that one in general
cannot assume such a steady state.  Beyond this, will the gas that we see at any
density at the current epoch turn into stars later?  Yet  
another conceptual problem is what volume(s)
should one choose to examine in order to determine a threshold for or efficiency of
making stars.  In the picture we have presented here, with continuous evolution
of lower-density gas into high-density, star-forming gas, it is not obvious how
one would choose any particular scale.  As stars can {\em only} be formed from
gas that has much higher densities than that of the interstellar medium in general, it is
not clear what one learns from connecting young stars with the very densest gas.  
Extragalactic observations over large scales
may sidestep these questions to some extent by averaging over sufficient volumes
that steady states may be achieved; but then the question is whether physical
insight is suppressed by such global averaging.  A fuller understanding of
star formation must involve the cycling of gas between low- and high-density phases.

\acknowledgments

AB acknowledges useful conversations with Mark Krumholz and Clare Dobbs, LH with
Javier Ballesteros-Paredes.  We also acknowledge a useful referee report from
Neal Evans.
This work was supported in part by NSF grant AST-0807305.

\end{document}